\documentstyle[twoside,psfig]{article}

\catcode`\@=11
\long\def\@makefntext#1{
\protect\noindent \hbox to 3.2pt {\hskip-.9pt  
$^{{\eightrm\@thefnmark}}$\hfil}#1\hfill}		%CAN BE USED 

\def\thefootnote{\fnsymbol{footnote}}
\def\@makefnmark{\hbox to 0pt{$^{\@thefnmark}$\hss}}	%ORIGINAL 
	
\def\ps@myheadings{\let\@mkboth\@gobbletwo
\def\@oddhead{\hbox{}
\rightmark\hfil\eightrm\thepage}   
\def\@oddfoot{}\def\@evenhead{\eightrm\thepage\hfil
\leftmark\hbox{}}\def\@evenfoot{}
\def\sectionmark##1{}\def\subsectionmark##1{}}

\oddsidemargin=\evensidemargin
\renewcommand{\thefootnote}{\fnsymbol{footnote}}

\newcounter{sectionc}\newcounter{subsectionc}\newcounter{subsubsectionc}
\renewcommand{\section}[1] {\vspace{12pt}\addtocounter{sectionc}{1} 
\setcounter{subsectionc}{0}\setcounter{subsubsectionc}{0}\noindent 
	{\tenbf\thesectionc. #1}\par\vspace{5pt}}
\renewcommand{\subsection}[1] {\vspace{12pt}\addtocounter{subsectionc}{1} 
	\setcounter{subsubsectionc}{0}\noindent 
	{\bf\thesectionc.\thesubsectionc. {\kern1pt \bfit #1}}\par\vspace{5pt}}
\renewcommand{\subsubsection}[1] {\vspace{12pt}\addtocounter{subsubsectionc}{1}
	\noindent{\tenrm\thesectionc.\thesubsectionc.\thesubsubsectionc.
	{\kern1pt \tenit #1}}\par\vspace{5pt}}

\newcounter{appendixc}
\newcounter{subappendixc}[appendixc]
\newcounter{subsubappendixc}[subappendixc]
\renewcommand{\thesubappendixc}{\Alph{appendixc}.\arabic{subappendixc}}
\renewcommand{\thesubsubappendixc}
	{\Alph{appendixc}.\arabic{subappendixc}.\arabic{subsubappendixc}}

\renewcommand{\appendix}[1] {\vspace{12pt}
        \refstepcounter{appendixc}
        \setcounter{figure}{0}
        \setcounter{table}{0}
        \setcounter{lemma}{0}
        \setcounter{theorem}{0}
        \setcounter{corollary}{0}
        \setcounter{definition}{0}
        \setcounter{equation}{0}
        \renewcommand{\thefigure}{\Alph{appendixc}.\arabic{figure}}
        \renewcommand{\thetable}{\Alph{appendixc}.\arabic{table}}
        \renewcommand{\theappendixc}{\Alph{appendixc}}
        \renewcommand{\thelemma}{\Alph{appendixc}.\arabic{lemma}}
        \renewcommand{\thetheorem}{\Alph{appendixc}.\arabic{theorem}}
        \renewcommand{\thedefinition}{\Alph{appendixc}.\arabic{definition}}
        \renewcommand{\thecorollary}{\Alph{appendixc}.\arabic{corollary}}
        \renewcommand{\theequation}{\Alph{appendixc}.\arabic{equation}}
        \noindent{\tenbf Appendix \theappendixc #1}\par\vspace{5pt}}
\newcommand{\subappendix}[1] {\vspace{12pt}
        \refstepcounter{subappendixc}
        \noindent{\bf Appendix \thesubappendixc. {\kern1pt \bfit #1}}
	\par\vspace{5pt}}
\newcommand{\subsubappendix}[1] {\vspace{12pt}
        \refstepcounter{subsubappendixc}
        \noindent{\rm Appendix \thesubsubappendixc. {\kern1pt \tenit #1}}
	\par\vspace{5pt}}

\newcommand{\textlineskip}{\baselineskip=13pt}
\newcommand{\smalllineskip}{\baselineskip=10pt}

\def\eightcirc{
\begin{picture}(0,0)
\put(4.4,1.8){\circle{6.5}}
\end{picture}}
\def\eightcopyright{\eightcirc\kern2.7pt\hbox{\eightrm c}} 

\newcommand{\copyrightheading}[1]
	{\vspace*{-2.5cm}\smalllineskip{\flushleft
	{\footnotesize Modern Physics Letters A, #1}\\
	{\footnotesize $\eightcopyright$\, World Scientific Publishing
	 Company}\\
	 }}

\newcommand{\publisher}[2]{{\begin{center}\footnotesize\smalllineskip 
	Received #1\\
	Revised #2
	\end{center}
	}}

\def\abstracts#1#2#3{{
	\centering{\begin{minipage}{4.5in}\footnotesize\baselineskip=10pt
	\parindent=0pt #1\par 
	\parindent=15pt #2\par
	\parindent=15pt #3
	\end{minipage}}\par}} 

\newcommand{\bibit}{\nineit}
\newcommand{\bibbf}{\ninebf}
\renewenvironment{thebibliography}[1]
	{\frenchspacing
	 \ninerm\baselineskip=11pt
	 \begin{list}{\arabic{enumi}.}
        {\usecounter{enumi}\setlength{\parsep}{0pt}     
	 \setlength{\leftmargin 12.7pt}{\rightmargin 0pt} %FOR 1--9 ITEMS
         \setlength{\itemsep}{0pt} \settowidth
	{\labelwidth}{#1.}\sloppy}}{\end{list}}

\newcounter{itemlistc}
\newcounter{romanlistc}
\newcounter{alphlistc}
\newcounter{arabiclistc}

\newcommand{\fcaption}[1]{
        \refstepcounter{figure}
        \setbox\@tempboxa = \hbox{\footnotesize Fig.~\thefigure. #1}
        \ifdim \wd\@tempboxa > 5in
           {\begin{center}
        \parbox{5in}{\footnotesize\smalllineskip Fig.~\thefigure. #1}
            \end{center}}
        \else
             {\begin{center}
             {\footnotesize Fig.~\thefigure. #1}
              \end{center}}
        \fi}

\newcommand{\tcaption}[1]{
        \refstepcounter{table}
        \setbox\@tempboxa = \hbox{\footnotesize Table~\thetable. #1}
        \ifdim \wd\@tempboxa > 5in
           {\begin{center}
        \parbox{5in}{\footnotesize\smalllineskip Table~\thetable. #1}
            \end{center}}
        \else
             {\begin{center}
             {\footnotesize Table~\thetable. #1}
              \end{center}}
        \fi}

\def\@citex[#1]#2{\if@filesw\immediate\write\@auxout
	{\string\citation{#2}}\fi
\def\@citea{}\@cite{\@for\@citeb:=#2\do
	{\@citea\def\@citea{,}\@ifundefined
	{b@\@citeb}{{\bf ?}\@warning
	{Citation `\@citeb' on page \thepage \space undefined}}
	{\csname b@\@citeb\endcsname}}}{#1}}

\newif\if@cghi
\def\cite{\@cghitrue\@ifnextchar [{\@tempswatrue
	\@citex}{\@tempswafalse\@citex[]}}
\def\citelow{\@cghifalse\@ifnextchar [{\@tempswatrue
	\@citex}{\@tempswafalse\@citex[]}}
\def\@cite#1#2{{$\null^{#1}$\if@tempswa\typeout
	{IJCGA warning: optional citation argument 
	ignored: `#2'} \fi}}

\def\pmb#1{\setbox0=\hbox{#1}
	\kern-.025em\copy0\kern-\wd0
	\kern.05em\copy0\kern-\wd0
	\kern-.025em\raise.0433em\box0}

\def\fnt#1#2{\footnotetext{\kern-.3em
	{$^{\mbox{\scriptsize #1}}$}{#2}}}

\def\fpage#1{\begingroup
\voffset=.3in
\thispagestyle{empty}\begin{table}[b]\centerline{\footnotesize #1}
	\end{table}\endgroup}

\def\runninghead#1#2{\pagestyle{myheadings}
\markboth{{\protect\footnotesize\it{\quad #1}}\hfill}
{\hfill{\protect\footnotesize\it{#2\quad}}}}
\font\tenrm=cmr10
\font\tenit=cmti10 
\font\tenbf=cmbx10
\font\bfit=cmbxti10 at 10pt
\font\ninerm=cmr9
\font\nineit=cmti9
\font\ninebf=cmbx9
\font\eightrm=cmr8

\def\qed{\hbox{${\vcenter{\vbox{			%HOLLOW SQUARE
   \hrule height 0.4pt\hbox{\vrule width 0.4pt height 6pt
   \kern5pt\vrule width 0.4pt}\hrule height 0.4pt}}}$}}

\renewcommand{\thefootnote}{\fnsymbol{footnote}}	%USE SYMBOLIC FOOTNOTE

\begin{document}
\setlength{\textheight}{7.7truein}  %for 2nd page onwards

\runninghead{Comment on ``Evidence for Neutrinoless Double Beta Decay"}
{Comment on ``Evidence for Neutrinoless Double Beta Decay"}

\normalsize\textlineskip
\thispagestyle{empty}
\setcounter{page}{1}

\copyrightheading{}			%{Vol. 0, No.0 (1992) 000--000}

\vspace*{0.88truein}
\fpage{1}
\centerline{\bf COMMENT ON ``EVIDENCE FOR NEUTRINOLESS}
\baselineskip=13pt
\centerline{\bf DOUBLE BETA DECAY"}
\vspace*{0.37truein}

\centerline{\footnotesize C. E. Aalseth$^1$, F. T. Avignone III$^2$, A. Barabash$^3$, F. Boehm$^4$, R. L. Brodzinski$^1$, J. I. Collar$^5$, }
\centerline{\footnotesize P. J. Doe$^6$, H. Ejiri$^7$, S. R. Elliott$^6$\footnote{Corresponding author}, E. Fiorini$^8$, R.J. Gaitskell$^9$, G. Gratta$^{10}$,}
\centerline{\footnotesize  R. Hazama$^6$, K. Kazkaz$^6$, G. S. King III$^2$, R. T. Kouzes$^1$, H. S. Miley$^1$, M. K. Moe$^{11}$, A. Morales$^{12}$, }
\centerline{\footnotesize   J. Morales$^{12}$, A. Piepke$^{13}$, R. G. H. Robertson$^6$, W. Tornow$^{14}$, P. Vogel$^4$, R. A. Warner$^1$,  J. F. Wilkerson$^6$}
\baselineskip=12pt

\centerline{\footnotesize\it $^1$Pacific Northwest National Laboratory, Richland, WA, 99352, USA }
\centerline{\footnotesize\it $^2$Department of Physics and Astronomy, University of South Carolina, Columbia, SC 29208, USA }
\centerline{\footnotesize\it $^3$Institute for Theoretical and Experimental Physics, Moscow 117259, Russia }
\centerline{\footnotesize\it $^4$Department of Physics, California Institute of Technology, Pasadena, CA 91125, USA}
\centerline{\footnotesize\it $^5$Enrico Fermi Institute, University of Chicago, Chicago, IL 60637, USA }
\centerline{\footnotesize\it $^6$Center for Experimental Nuclear Physics and Astrophysics, University of Washington, Seattle, WA 98195, USA }
\centerline{\footnotesize\it $^7$International Institute for Advanced Studies, Kizu-cho, Kyoto, 619-0025 and JASRI, SPring-8, Hyogo 679-5198, Japan }
\centerline{\footnotesize\it $^7$Emeritus, Research Center for Nuclear Physics, Osaka University, Ibaraki, Osaka 567, Japan}
\centerline{\footnotesize\it $^8$Dipartimento di Fisica dell' Universita' di Milano-Bicocca and Istituto Nazionale di Fisica Nucleare, Sezione di Milano, Italy }
\centerline{\footnotesize\it $^9$Department of Physics, Brown University, Providence, RI 02912, USA }
\centerline{\footnotesize\it $^{10}$Physics Department, Stanford University, Stanford, CA 94305, USA }
\centerline{\footnotesize\it $^{11}$Emeritus, Department of Physics and Astronomy, University of California, Irvine, CA 92697, USA } 
\centerline{\footnotesize\it $^{12}$Laboratory of Nuclear and High Energy Physics, University of Zaragoza, 50009 Zaragoza, Spain}
\centerline{\footnotesize\it $^{13}$Department of Physics and Astronomy, University of Alabama, Tuscaloosa, AL 35487, USA }
\centerline{\footnotesize\it $^{14}$Department of Physics, Duke University, Durham, NC 27708, USA }
\centerline{\footnotesize\it }
\vspace*{10pt}
\baselineskip=10pt

\vspace*{0.225truein}

\publisher{(received date)}{(revised date)}

\vspace*{0.21truein}
\abstracts{We comment on the recent claim for the experimental observation of neutrinoless double-beta 
decay. We discuss several limitations in the analysis provided
in that paper and conclude that there is no basis for the presented claim.}{}{}

%\vspace*{10pt}
%\keywords{The contents of the keywords}

%\textlineskip			%) USE THIS MEASUREMENT WHEN THERE IS
%\vspace*{12pt}			%) NO SECTION HEADING

\vspace*{1pt}\textlineskip	%) USE THIS MEASUREMENT WHEN THERE IS
\section{Introduction}	%) A SECTION HEADING
\vspace*{-0.5pt}
\noindent
In a paper by Klapdor-Kleingrothaus, Dietz, Harney, and Krivosheina\cite{KDHK} (Hereafter referred to
as KDHK)  evidence is claimed for zero-neutrino double-beta decay in $^{76}$Ge. The high quality 
data, upon which this claim is based, was compiled by the careful efforts of the Heidelberg-Moscow collaboration,
and is well documented\cite{HM}. However, the analysis in KDHK makes an extraordinary claim, and therefore requires
very solid substantiation. In this letter, we outline our concerns for the claim of evidence.

Unfortunately, a large number of issues were not addressed in KDHK. Some of these are:

\begin{enumerate}

 \item There is no null hypothesis analysis demonstrating that the data require a peak.
Furthermore, no simulation has been presented to demonstrate that the analysis correctly finds true peaks or
that it would find no peaks if none existed. Monte Carlo simulations of spectra containing different 
numbers of peaks are needed to confirm the significance of any found peaks.

 \item There are three unidentified peaks in the region of analysis that 
have greater significance than the 2039-keV peak. There is no discussion of the
origin of these peaks.

 \item There is no 
discussion of how sensitive the conclusions are to different mathematical models. 
There is a previous Heidelberg-Moscow publication\cite{HM}  that gives a lower limit of 1.9 $\times$ 10$^{25}$ y (90\% confidence level).
This is in conflict with the ``best value" of the new KDHK paper of 1.5 $\times$ 10$^{25}$ y.  This indicates
a dependence of the results on the analysis model and the
background evaluation.  

\end{enumerate}

A number of other cross checks of the result should also be performed. For example,
 there is no discussion of how a variation of the size of the chosen analysis
window affects the significance of the hypothetical peak. There is no relative peak strength 
analysis of all the $^{214}$Bi peaks. Quantitative evaluations should be made
on the four $^{214}$Bi peaks in the region of interest. There is no statement of the net 
count rate of the peaks other than the 2039-keV peak. There is no presentation of the entire 
spectrum. As a result, it is difficult to compare relative strengths of peaks.
 There is no discussion of the relative peak strengths before and after the single-site-event 
cut. This is needed to elucidate the model for the origin of the peaks.

Our investigations of several of these issues create doubt as to 
the validity of the paper's conclusions. Below, we demonstrate this by briefly
discussing two items. Our analyses suggest that,
 at best, KDHK failed in presenting a strong case for their extraordinary claim, and  
at worst, that their analyses or assumptions have led to an incorrect claim.

\setcounter{footnote}{0}
\renewcommand{\thefootnote}{\alph{footnote}}

\section{The Choice of Window and Background}
\noindent
In \frenchspacing{Fig. 2} of KDHK, the count rate in the region of 2000-2080 keV 
 is 0.168 c/(keV$\cdot$kg$\cdot$yr), and in the 2034-2045 keV region it is 0.167 c/(keV$\cdot$kg$\cdot$yr).
These numbers indicate a flat background and very little signal. However, KDHK found numerous peaks in the
2000-2080 keV region in their search for a peak. Next, they constrained their double-beta decay ($\beta\beta$) analysis 
to a small region that excluded these
peaks. An analysis only within that limited region is used to claim a 2039-keV peak at the 2-3 $\sigma$ level.
 The conclusion in KDHK must depend on the choice of window and on the number of peaks 
in the region near the window. In particular, it requires
that the other peaks found are real so that the background level can be diminished.

\section{The Relative Strength of the $^{214}$Bi Peaks}
\noindent
The KDHK paper provides  data only in the 2000-2080 keV region. However a recent
paper by the Heidelberg-Moscow 
collaboration\cite{HM} (hereafter referred to as HM) shows the entire spectrum\footnote{\protect 
Fig. 1 in HM has an incorrect caption and the Figure is actually for only 1 detector.
Therefore we multiplied the net peak counts from that detector by 4 so it would correspond to a spectrum
 composed of 4 detectors and could be compared to Fig. 2 in KDHK. A similar analysis
using a 49.03 kg-y, 5-detector spectrum from Ref. \cite{Baudis} leads to 
the same qualitative conclusions.}
 ~from 100-2700 keV, which permits
a relative-intensity peak-analysis. Although the two data sets are not entirely congruent, they are very similar, having 47.4 kg-y of 
data (HM) and 46.5 kg-y of data (KDHK). They both quote comparable
backgrounds in the 2000-2080 keV region.  

The table below summarizes an analysis of the $^{214}$Bi peak intensities,
as given in HM and in KDHK. The first column gives the peak energy of
 seven of the $^{214}$Bi lines: three major
 lines spanning the region of interest and the four weak lines that can appear  
in the region of interest. KDHK claims to have observed lines at the positions of the 4 weak $^{214}$Bi lines.

\begin{table}[htbp]
\tcaption{A comparison of the intensities of the $^{214}$Bi lines. The count rates for the peaks labeled
as {\it Ref. Peak} come from Ref.\cite{HM}. The relative efficiency for the peaks in the 2000-2080 keV region
is an interpolated value based on the 3 reference peaks.}
\centerline{\footnotesize\smalllineskip
\begin{tabular}{rccccc}\\
\hline
Peak    &   Rate        	   &  Branching     	&  Relative	               &  Expected Rate     \\
 (keV)  & (c/(kg$\cdot$yr)) & Ratio\cite{TOI} &  Efficiency	             &    (c/(kg$\cdot$yr))       \\
\hline 
609.3	  &     44           	&   44.8\%        &  	 1	                    &  Ref. Peak             \\
1764.5	 &      16          	&   15.36\%  	    &  1.08		                  &  Ref. Peak            \\
2010.7	 &     	-            &  	0.05\%        &  1.11                    & 	0.05	              \\
2016.7	 &     	-            &  	0.0058\%     	&  1.11	                   &  0.006	              \\
2021.8	 &     	-            &  	0.02\%        &  1.11	                   &  0.02	             \\
2052.9	 &     	-            &  	0.078\%       &  1.11	                   &  0.08	              \\
2204.2	 &     	5.2 	        &   4.86\%       	&  1.13		                  &  Ref. Peak             \\
\hline \\
\end{tabular}}
\end{table}

The rates for the 3 major peaks given in Column 2 were calculated by integrating the spectrum in HM.
For those 3 peaks, the fourth column is the relative peak count rate divided by the branching 
fraction.  If the efficiency were the same for all peak energies, these should
 all have the same value.  In fact, peak efficiency is rarely independent of
energy, and here it actually increases slightly with
 increasing energy.  

Since the relative counting efficiency is virtually flat as a function of energy, all peaks in the 2000-2080 keV region
are assigned the same interpolated number. The last column uses the 
measured $^{214}$Bi disintegration rate from the three major (and prominent) peaks
 and calculates the expected count rate for the four minor $^{214}$Bi peaks in KDHK.
 These rates are all $\leq$ 0.08 c/(kg$\cdot$yr). For a peak width of $\approx$4 keV, this corresponds to 0.02 c/(keV$\cdot$kg$\cdot$yr)
and therefore too
low to be observed as peaks superimposed upon a background of 0.17 c/(keV$\cdot$kg$\cdot$yr) as shown in Fig. 2 of KDHK.
The strongest of these peaks would produce only
about 4 counts in 46.5 kg-y over an expected background of $\approx$32 counts. One is led to conclude that there
are no observable $^{214}$Bi peaks in the 2000-2080 keV region. That is, the $^{214}$Bi peaks found by KDHK in the
 region of interest appear to be spurious.

\section{Conclusions}
\noindent
A simple
analysis of the $^{214}$Bi peaks suggests that the peak finding procedure used by KDHK can produce 
spurious peaks near the $\beta\beta(0\nu)$ endpoint. However, the existence of these claimed peaks is
crucial to the KDHK claim of a peak at 2039 keV, which is interpreted as zero-neutrino double-beta decay.
Hence, all the peaks claimed in the 2000-2080 keV region may be spurious, and 
the entire count rate therefore attributable to a flat uniform background. Alternatively, if all the 
peaks are real but unidentified, the putative 2039-keV feature may be simply
 another of those unidentified lines.

These two examples emphasize the importance of addressing
all the items listed in the Introduction.
 By failing to address these issues, the KDHK paper does not provide sufficient
 support for its claim of evidence for $\beta\beta(0\nu)$.

\section{Acknowledgements}
\noindent
We would like to acknowledge supportive discussions with  
Dale Anderson, Laura Baudis, Theodore Bowyer,  David Jordan, Bela Majorovits, 
 William Pitts, Eric Smith, Bob Thompson, Albert Young, and
Igor Kirpichnikov.

The preprint by Feruglio, Strumia, and Vissani\cite{FSV} with independent criticisms of the KDHK result
became available during the final preparation of this manuscript.

\end{document}